\documentclass[reprint, amsmath,amssymb,aps]{revtex4-2}

\usepackage{graphicx}
\usepackage{dcolumn}
\usepackage{bm}
\usepackage{physics}
\usepackage{amsmath}
\usepackage{amssymb}

\begin{document}

\preprint{APS/123-QED}

\title{Zeno Effect Suppression of Gauge Drift in Quantum Simulations}

\author{Carter Ball}
\email{cball12@umd.edu}
\affiliation{Department of Physics, University of Maryland, College Park, MD 20742, USA}
\author{Thomas D. Cohen}
\email{cohen@umd.edu}
\affiliation{Department of Physics, University of Maryland, College Park, MD 20742, USA}

\date{\today}

\begin{abstract}
 Quantum simulation of lattice gauge theories is a promising tool for the study of many complicated problems including ones with real-time dynamics. For gauge theories, however, there is a major challenge in maintaining gauge invariance during time evolution. Such theories have a full Hilbert space that is larger than the physical space---the set of states which are gauge invariant or equivalently respect the Gauss law.  While an exact implementation of Hamiltonian dynamics starting in the physical Hilbert space will keep the system in the physical space, various types of errors will inevitably produce components outside of it. This work proposes a method of suppressing this gauge drift via the Zeno effect. As in the standard picture of the Zeno effect, our method relies on frequent projection onto the physical subspace. Additionally, a technique is discussed to reduce the speed of the gauge drift, which helps to reduce the required frequency of projections. We demonstrate our method on a $\mathbb{Z}_2$ gauge theory toy model. 
\end{abstract}

\maketitle

\section{\label{sec:level1}Introduction}
Quantum computers promise to be vital tools for the simulation of many fundamental yet complicated problems \cite{QuantCompSim1,QuantCompSim2,QuantCompSim3,QuantCompSim4,QuantCompSim5} from a range of subfields, including particle physics, nuclear physics, and condensed matter. The simulation and study of complex systems often requires much more computing power than classical computers can provide \cite{ComputePower1,ComputePower2}; thus, there is a present need to map these systems onto quantum devices. Analog quantum devices manufacture a system that can imitate, at least in some regime, the desired theory while digital quantum devices map a theory onto an array of qubits and control it via quantum gates \cite{DigAnal1}. Much research has gone into studying these quantum devices and how they can be used for quantum simulation \cite{QuantSim}. 

Like many complicated theories, lattice gauge theories can, in principle, benefit enormously from efficient and reliable quantum simulation \cite{LGTsim0,LGTsim1,LGTsim2,DigQuantSim1,DigQuantSim2,LiveWithIt, EnergyPenalty3,QuantSimUltracold,LGTsim3,LGTsim4,LGTsim5,LGTsim6,LGTsim7,LGTsim8,LGTsim9,LGTsim10}. These theories typically place matter degrees of freedom on the sites of the lattice while gauge field operators are placed on the links; it is typical to work in a Hamiltonian formulation with discretized space but continuous time \cite{LGTintro,HamiltonianLGT,LGToverview}. This is usually done in the temporal gauge \cite{TemporalGauge} (which sets the time component of gauge fields to zero) and by taking the continuous time limit. Note that the temporal gauge does not fully exhaust the gauge freedom, leaving a residual gauge freedom for the spatial components. Transformations of this residual gauge are generated by Gauss law operators (in analogy to classical electrodynamics). Physical states are then defined as those that are invariant under all residual gauge transformations. 

Much work has gone into the study of both Abelian and non-Abelian lattice gauge theories. Notably, maintaining non-Abelian gauge invariance is particularly difficult. When it comes to analog quantum simulation, many methods have been developed to address this gauge invariance problem, including constructing a system that makes gauge invaraince a consequence of internal symmetry such as angular momentum \cite{InternalSym1, InternalSym2}. Another analog simulation method is to impose an energy penalty on the unphysical states \cite{EnergyPenalty1,EnergyPenalty2,EnergyPenalty3,EnergyPenalty4,EnergyPenalty5,EnergyPenalty6,EnergyPenalty7,EnergyPenalty8,EnergyPenalty9,EnergyPenalty10}, often by adding a term to the Hamiltonian proportional to the square of the Gauss' law operators. Then, the lowest energy levels of the system are the physical ones, so keeping the system in the low energy regime results, to good approximation, in keeping the system within the physical subspace of the much larger Hilbert space. Variations on this method have been developed, including one where the addition to the Hamiltonian is directly proportional to the Gauss' law operators \cite{SingleBodyEnergyPenalty}. Another such variation on the energy penalty method uses local pseudogenerators, which exactly match the Gauss' law operators in the relevant physical subspace but not in unphysical sectors, allowing these operators to be much less complex and thus cheaper to engineer in practice \cite{LocalPseudoGens1,LocalPseudoGens2,LocalPseudoGens3}. Interestingly, for sufficiently large energy penalties, the protection term added to the Hamiltonian, constructed with local pseudogenerators, can be seen as a strong projector which constrain the dynamics of the system to quantum Zeno subspaces within a given timescale \cite{LocalPseudoGens3,QZsubspaces}.

The method presented in this paper concerns digital quantum simulation. Ideally, any method to simulate a non-Abelian gauge theory of interest, such as quantum chromodynamics (QCD), would only simulate the gauge invariant (i.e. physical) states. Methods of this kind have been proposed which solve the Gauss law, either eliminating the gauge fields \cite{Encode1,Encode2,Encode3} or the matter fields \cite{Encode4,Encode5,Encode6,Encode7,Encode8}. Another method uses dual formulations of the gauge theories to enforce the Gauss law \cite{Dual1,Dual2,Dual3}. While the methods discussed above don't have to deal with unphysical states, their physical implementation may be more challenging. 

Thus, many methods simulate all states in the Hilbert space of the system, and must contend with keeping the system within the exponentially smaller physical subspace. Crucially, work has been done to develop methods of time evolution that respect gauge invariance \cite{DigQuantSim1, YukariDigQuantSim}. Thus, ideally time evolving an initial state that is constructed to be physical would keep the system within the physical subspace. However, quantum noise, gate errors, and various errors from approximations can lead to the system picking up contributions from unphysical states. Previous work has addressed this problem in a variety of ways. Perhaps the most straightforward is to simply accept the unphysical contributions on the premise that these contributions are reduced as the time step is decreased \cite{LiveWithIt}. Another method is dynamical decoupling \cite{DynamicalDecoupling}, a concept from quantum control theory where local symmetries are maintained via periodic driving. Also, oracles have been developed \cite{AbelianOracle} to check for violations of Gauss' law for Abelian lattice gauge theories, with similar oracles for non-Abelian theories under development. 

Furthermore, a method of engineering classical noise to constrain the system to the physical subspace \cite{ZenoRandomNoise} has been proposed. This method serves as a classical analog to the quantum Zeno effect by using classical fluctuations of a perturbing field to dissipate unwanted evolution into unphysical regions of the Hilbert space. The method of this paper, in contrast, rests on the standard quantum Zeno effect \cite{Zeno}, which uses frequent projections to keep the system within a desired space; for our purposes the desired space is of course the physical subspace of the system's Hilbert space. 

This paper is organized as follows: section II outlines our method, which takes inspiration from the rodeo algorithm\cite{Rodeo1}---a method of projection---and a technique of unitary gauge drift suppression. Section III demonstrates our method on a toy model of simulating pure $\mathbb{Z}_2$ gauge theory on a 2x2 lattice with periodic boundary conditions. Section IV goes into more detail about the implementation of our method, and Section V provides concluding thoughts and discusses the required future work. 
\section{Method}
The method exploits the quantum Zeno effect to maintain a close approximation of gauge invariance. For a non-Abelian theory, the gauge invariant, or physical, states are given by those states $\ket{\psi}$ that satisfy
\begin{align}
    G^a_x\ket{\psi} = 0
\end{align}
for all $a$ and $x$, where $G^a_x$ is the $a$th Gauss' law operator at site $x$. Note that for compact Lie gauge groups (e.g. $\mathit{SU(N)}$), these $G^a_x$ operators generate local gauge transformations.

To leverage the Zeno effect, we must invoke a method of projection onto the physical subspace. Our method adapts previous work on the rodeo algorithm \cite{Rodeo1,Rodeo2,Rodeo3,RodeoOpt}. We also incorporate a method of suppressing coherent gauge drift, discussed below. 
\subsection{Projection via the Rodeo Algorithm}
\begin{figure}[h]
    \centering
    \includegraphics[width=\linewidth]{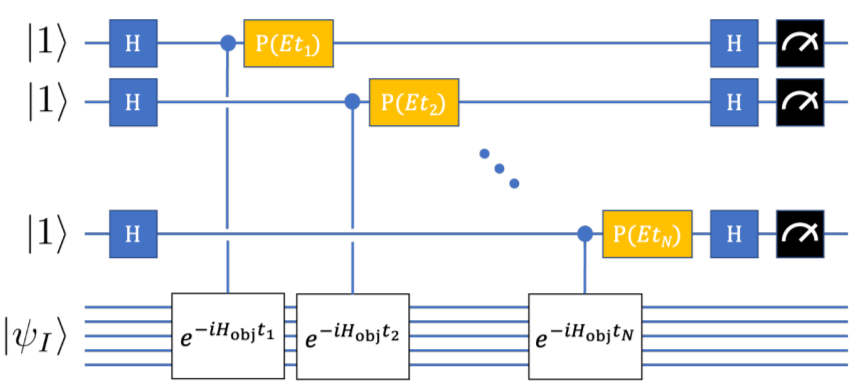}
    \caption{The circuit diagram for the rodeo algorithm \cite{Rodeo1}}
    \label{fig:Rodeo}
\end{figure}

The rodeo algorithm was developed to prepare eigenvectors of a given Hamiltonian. See figure \ref{fig:Rodeo} for the circuit diagram of the algorithm \cite{Rodeo1}. 

The rodeo algorithm aims to construct, to good approximation, an eigenstate of a system with a chosen energy by "shaking off" amplitudes of other, unwanted eigenstates, like a bull shaking off its rider at a rodeo. The algorithm accomplishes this task with ancillary qubits as shown in the circuit diagram of Fig.~\ref{fig:Rodeo}. To see how the rodeo algorithm constructs the desired eigenstate, we will write out how it works for one ancillary qubit. Starting with the system in the initial state $\ket{\psi_I}$ and the ancillary qubit initialized to $\ket{1}$, we have
\begin{align}
    \begin{bmatrix}
        \frac{I}{\sqrt{2}} & \frac{I}{\sqrt{2}} \\
        \frac{I}{\sqrt{2}} & -\frac{I}{\sqrt{2}}
    \end{bmatrix}
    \begin{bmatrix}
        I & 0 \\
        0 & Ie^{iEt}
    \end{bmatrix}
    \begin{bmatrix}
        I & 0 \\
        0 & e^{-iH_{obj}t}
    \end{bmatrix}
    \begin{bmatrix}
        \frac{I}{\sqrt{2}} & \frac{I}{\sqrt{2}} \\
        \frac{I}{\sqrt{2}} & -\frac{I}{\sqrt{2}}
    \end{bmatrix}
    \begin{bmatrix}
        0 \\
        \ket{\psi_I}
    \end{bmatrix} = \\ 
    \begin{bmatrix}
        \frac{I}{\sqrt{2}} & \frac{I}{\sqrt{2}} \\
        \frac{I}{\sqrt{2}} & -\frac{I}{\sqrt{2}}
    \end{bmatrix}
    \begin{bmatrix}
        I & 0 \\
        0 & Ie^{iEt}
    \end{bmatrix}
    \begin{bmatrix}
        I & 0 \\
        0 & e^{-iH_{obj}t}
    \end{bmatrix}
    \begin{bmatrix}
        \frac{1}{\sqrt{2}}\ket{\psi_I} \\
        -\frac{1}{\sqrt{2}}\ket{\psi_I}
    \end{bmatrix} = \\
    \begin{bmatrix}
        \frac{I}{\sqrt{2}} & \frac{I}{\sqrt{2}} \\
        \frac{I}{\sqrt{2}} & -\frac{I}{\sqrt{2}}
    \end{bmatrix}
    \begin{bmatrix}
        \frac{1}{\sqrt{2}}\ket{\psi_I} \\
        -\frac{1}{\sqrt{2}}e^{-i(H_{obj}-E)t}\ket{\psi_I}
    \end{bmatrix} = \\
    \begin{bmatrix}
        \frac{1}{2}(I - e^{-i(H_{obj}-E)t})\ket{\psi_I} \\
        \frac{1}{2}(I + e^{-i(H_{obj}-E)t})\ket{\psi_I}
    \end{bmatrix}
\end{align}
where above, $H_{obj}$ is the Hamiltonian of the system of interest (ie the object system) and $E$ is the desired energy of the constructed state. Equation (2) writes out the actions of (reading from right to left) the Hadamard, controlled time evolution, phase shift, and Hadamard gates, respectively, on the initial state. Note that the first Hadamard gate gives equal amplitude, up to a sign, to the up (ie $\ket{0}$) and down (ie $\ket{1}$) components of the ancillary qubit (equation (3)), then the controlled time evolution and phase shift gates give a phase only to the down component (equation (4)), and finally the last Hadamard gate causes interference between the up and down components (equation (5)). 

The final step in the rodeo algorithm is to measure the ancillary qubit(s). A successful implementation of the rodeo algorithm measures every ancillary qubit in the state $\ket{1}$ at the end. In the single ancillary qubit example above, this results in the transformation 
\begin{align}
    \ket{\psi_I} \rightarrow \frac{1}{2}(I + e^{-i(H_{obj}-E)t})\ket{\psi_I}
\end{align}
To see the benefits of this transformation, we decompose our initial state into energy eigenstates $\ket{\psi_I} = \sum\limits_{n}c_n\ket{n}$ where $H_{obj}\ket{n} = E_n\ket{n}$. The amplitude of each energy eigenstate then transforms as
\begin{align}
    c_n \rightarrow \frac{1}{2}c_n(1+e^{-i(E_n-E)t})
\end{align}
which means the probability of each energy eigenstate transforms as 
\begin{align}
    |c_n|^2 &\rightarrow |c_n|^2|\frac{1}{2}(1+e^{-i(E_n-E)t})|^2 \\
    &\rightarrow |c_n|^2\cos^2((E_n-E)\frac{t}{2})
\end{align}
Thus, the probability for the system to be in a particular energy eigenstate $E_{n}$ is multiplied by a factor $\cos^2((E_{n} - E)\frac{t}{2})$, which is notably 1 if $E = E_{n}$ and less than 1 otherwise. This is how the rodeo algorithm works: it constructs an approximate version of the state of energy $E$ by suppressing all other states with a $\cos^2(\frac{1}{2}t\Delta E)$ factor. Stronger suppression of contributions from unwanted energy eigenstates can be accomplished with multiple rounds of this algorithm, as discussed in \cite{Rodeo1,Rodeo2,Rodeo3}. Note that in order to implement a round of the rodeo algorithm, a time $t$ must be chosen. If the times are chosen strategically for the various rounds, remarkably strong suppression can be achieved with very modest resources \cite{RodeoOpt}.  

We note here that the rodeo algorithm will fail to project onto the state of interest with finite probability (i.e. if any ancillary qubit is not measured in $\ket{1}$), where failure means that the resulting state is orthogonal to the target state. Crucially, we note that the probability of a successful projection is proportional to the overlap of the initial state of the system with the target state of the projection. As we plan to use this projection to leverage the Zeno effect, which means making sufficiently many measurements so that the overlap with the target subspace is close to unity, the rate of failure can be made negligibly small. The chance of failure serves as the probabilistically defined time span for which our method works; for example, if the probability for a failed projection is one in a million, we would expect one failure within one million successive projections.

So far, we have focused on the rodeo algorithm due to its key feature of leaving desired amplitudes (ie the amplitude of the desired energy eigenstate) untouched while suppressing all other amplitudes. Here, we note that there is no reason this algorithm works only for the construction of the eigenstates of the Hamiltonian. Thus, going forward, we will refer to rodeo projections as any projection of the form expressed in Fig.~\ref{fig:Rodeo} with some hermitian operator in place of the Hamiltonian. 

Recall that our goal is to use frequent projections onto the physical subspace to maintain gauge invariance of the system via the Zeno effect. With this in mind, we now discuss the form of a rodeo projection that we will use to project onto the physical subspace. We replace the Hamiltonian with a new hermitian operator that we will call $G^2$:
\begin{align}
    G^2 = \sum\limits_{a,x}(G^a_x)^2
\end{align}
From the above definition of the Gauss' law operators, we see that all physical states are eigenstates of $G^2$ with eigenvalue 0. Furthermore, all the other eigenstates will be unphysical and with eigenvalues greater than 0, since $G^2$ is positive definite by construction. It is this spectrum of $G^2$ that we will exploit when using rodeo projections: instead of constructing some eigenstate of a given Hamiltonian, we will construct a physical state by constructing a ground state of $G^2$ via rodeo projections. This means replacing $H_{obj}$ in Fig.~\ref{fig:Rodeo} with $G^2$; furthermore, as the eigenvalue of the desired eigenstate (ie a physical state) is 0, we will remove the phase shift gates, since the required phase shift is $0*t = 0$. 

This is the core of the method outlined in this paper: a rodeo projection with this $G^2$ operator will leave all physical amplitudes untouched while suppressing all unphysical amplitudes. Doing this projection frequently enough will keep the system in a close approximation of a gauge invariant state via the Zeno effect. 

There is one minor detail that we must clarify before moving on. The effectiveness of this projection is helped if the spectrum of the operator, in our case $G^2$, is discrete. That way there will be less contamination from the eigenstates nearest to the target eigenstate. Luckily, the spectrum of $G^2$ for our principal case of interest $SU(3)$ gauge theories, is in fact discrete. %This can be seen rather straightforwardly, by considering the three well-known $SU(2)$ subgroups, often associated with I-spin, U-spin, and V-spin. Note that every component $G^a$ is part of one of these subgroups (if in linear combination with other components). Now, introductory quantum mechanics \cite{QuantMech} provides a basic argument as to why the operators of an $SU(2)$ group have discrete spectra.
This can be seen very straightforwardly using basic quantum mechanics, though We will point out one small detail: the basic argument implicitly relies on the fact that all states have non-negative norms. While this is obviously true for physical states it is not necessarily true for the full Hilbert space of QCD.  However, we are working within a Hamiltonian formulation and using the temporal gauge, which is a ghostless gauge \cite{Ghostless}, meaning that we do in fact have non-negative norms for all states. Thus, all components $G^a$ have discrete spectra, so $G^2$ has a discrete spectrum as well. This means the ground state of $G^2$ is gapped, which is helpful when it comes to projection.

\subsection{Suppressing Coherent Gauge Drift}
The basic idea of the Zeno effect is that if you project back to the desired space before straying too far out of it (i.e. frequently enough), then the system stays in the desired space with high probability. This is for a simple reason: projections depend on the probability that the system is in the desired state, which corresponds to the square of the magnitude of the amplitude. Thus, if one projects when the amplitude for the system drifting into an unwanted state is some small value $\epsilon$, the probability of failure is $\epsilon^2$---a much smaller value.  

Consider what this means in practice if one wishes to study a system's time evolution over a time, $T$, and gauge-invariance-violating errors in the amplitude accrue at a rate $r$.  Suppose one makes $N$ evenly spaced projections (which for the sake of this analysis are assumed to be perfect) during the time evolution. The probability of a failure---obtaining a state outside the physical space---at each measurement is $(r T/N)^2$, and the probability of a successful run for the full time evolution (i.e. the probability of $N$ successes and $0$ failures) is given by
\begin{equation}
P_s = \left (1-\frac{r^2T^2}{N^2}  \right)^N \approx 1 - \frac{r^2T^2}{N}
\end{equation}
where the approximate equality holds in the limit of large $N$ and $1-P_s\ll 1$.  The key point is that for $N\gtrsim r^2 T^2$ there is a substantial probability that the full time evolution can occur with the system restricted to the physical space. Note that this means that by making sufficient measurements, the evolution can stay in the physical space even if the evolution time is much larger than $1/r$. This a major improvement.

Of course, if one makes a finite number of
projections there is a finite probability of failure---obtaining a state outside the physical space.  If that happens, the only recourse is to stop the calculation and begin again---an expensive prospect.  Thus, there is an optimization problem in choosing $N$: if $N$ is too large,  there is the computational cost of making more projections than are necessary, whereas if $N$ is too small than the failure rate is too high requiring the additional cost of starting over. The optimal value depends on the computation costs of the projection as well as the costs of running the simulation.

With this in mind, the method has been enlarged to include an additional way to suppress gauge drift, an approach developed in ref.~\cite{LLY}. we show that a mixed approach using both quantum Zeno and the the approach of ref.~\cite{LLY} is likely to be more efficient than either one separately.

The method of ref. ~\cite{LLY} is essentially an alternative version of the energy penalty method for suppressing coherent gauge drift. Here, we define coherent gauge drift as a unitary process $U_D$ that transforms a physical state $\ket{\psi}$ to
\begin{align} \label{eq:drift}
    U_D\ket{\psi} = \sqrt{1-\epsilon^2}\ket{\psi} + \epsilon\ket{\omega}
\end{align}
where $\ket{\omega}$ is a nonphysical state and $\epsilon$ characterizes the strength of the drift. Recall that an energy penalty method constructs a Hamiltonian $H_G$ with the behavior that $H_G\ket{\psi} = 0$ for physical states $\ket{\psi}$ and $H_G\ket{\omega} > 0$ for unphysical states $\ket{\omega}$. Adding $\lambda_{G}H_G$ for $\lambda_G \gg 0$ to the Hamiltonian of the system suppresses transitions to unphysical states via an energy penalty. The authors of \cite{LLY} note that in principle, building $H_G$ out of the fundamental fields of the lattice gauge theory would be quite difficult for a non-Abelian theory; however, quantum simulation via Trotterized time steps depends upon $e^{-iH_Gt}$ not solely $H_G$. Furthermore, they point out that this operator basically performs a random gauge transformation on the state of the system, which leaves physical states unchanged while unphysical states take on a phase. Thus, this approach dispenses with the difficult task of building $H_G$ and instead simply implements the principal effect of including $H_G$ via a random gauge transformation. In summary, the method of \cite{LLY} is to implement, after every Trotterization time step, a random gauge transformation. 

Our method aims to leverage this method wholesale: frequent projections back to the physical subspace is expensive, so we include a random gauge transformation after each Trotterization time step. This reduces the gauge drift per time step, allowing us to do projections to the physical subspace less frequently while still staying close enough to the physical subspace to leverage the quantum Zeno effect, which allows the suppression of whatever gauge drift that survives the random gauge transformations.

\subsection{Calibration}
Our method, as outlined above, is to conduct a random gauge transformation after every Trotterization time step as well as to conduct  projections in the form discussed above to project back to the physical subspace. One important detail that needs to  be specified is the frequency of projections. To this end, we include a calibration phase in the method---a set of initial studies---to best estimate the frequency of projections that constitutes a useful compromise between the minimization of gauge drift and the minimization of the number of projections. We note up front that in principle doing a projection after every time step would be most effective in terms of reducing gauge drift as the probability of failure of the projection is very low. The  problem with this is simply that it is extremely expensive computationally. In the next section we will outline results from a toy model that show that other, less expensive forms of the method work well. This suggests that realistic calculations can balance the resource costs with the requirements on effective gauge drift suppression.

We propose an approach that should be useful in situations when a simulation needs to be done repeatedly---in that case, a simple calibration phase to optimize the calculation is done at the outset: start with an initial trial frequency of projections based on whatever information and intuition one has about the system. Here, any previous knowledge about the system being simulated should be exploited, especially concerning the speed of gauge drift. With an initial frequency chosen, the time evolution of the system is run a small number of times with the chosen trial frequency of projections to see how long before the projection step fails. If it appears that failure will in general occur before the desired length of time of the simulation, the frequency of projections can be increased. This process can be repeated until an acceptable balance is found. 

Behind this simple calibration phase is the fact that the average gauge drift is linearly proportional to the number of time steps between projections. Intuitively, this also means the probability of a failed projection is linearly proportional to the number of time steps between projections. With these relations in mind, it is simple to tune the frequency of projections given a relatively few trial runs.

\section{Performance}
\begin{figure}[h]
    \centering
    \includegraphics[width=0.5\linewidth]{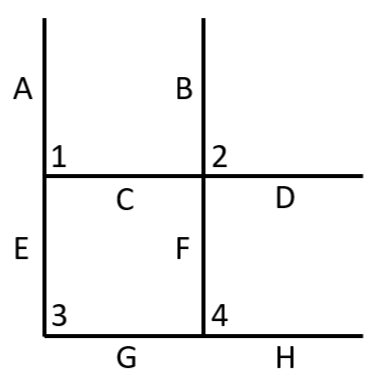}
    \caption{The diagram of the toy model with 4 sites and 8 links}
    \label{fig:Lattice}
\end{figure}
To study our algorithm, we consider a  small ``toy'' model.  It is a pure $\mathbb{Z}_2$ gauge theory in 2 spatial dimensions \cite{Z2}. We simulate this on a 2x2 lattice with periodic boundary conditions (see Figure \ref{fig:Lattice}). The purpose of using such a small simulation is simply that it allows one to explore the method with very little cost. The Hamiltonian for the toy model is 
\begin{align}
    H =  -g\sum\limits_{i=A}^H \sigma^x_i -k\sum\limits_{j=i}^4\square_j
\end{align}
where $g$ and $k$ are parameters, $\sigma^x_i$ is the Pauli x operator on the $i$th link, and $\square_j$ is the $j$th plaquette operator (where the $j$th plaquette is defined as the plaquette with the $i$th site in the bottom left corner). A plaquette operator in this context is a product of the $\sigma^z$ operators living on the four sides of a plaquette in the lattice; for example, the first plaquette operator is
\begin{align}
    \square_1 = \sigma^z_A\sigma^z_B\sigma^z_C\sigma^z_G
\end{align}

In this system, each site has an associated local gauge transformation $g_i$, constructed by taking the product of the $\sigma^x$ operators living on the four links emanating from the $i$th site on the lattice:
\begin{align}
    g_1 &= \sigma^x_A\sigma^x_C\sigma^x_D\sigma^x_E \\
    g_2 &= \sigma^x_B\sigma^x_C\sigma^x_D\sigma^x_F \\
    g_3 &= \sigma^x_A\sigma^x_E\sigma^x_G\sigma^x_H \\
    g_4 &= \sigma^x_B\sigma^x_F\sigma^x_G\sigma^x_H 
\end{align}
We note here that our gauge group $\mathbb{Z}_2$ is discrete, which means we cannot formulate Gauss' law via generators of gauge transformations. Instead, we note that the Hilbert space of the system is split into sectors defined by the eigenvalues of the gauge transformations in equations (15)-(18). Note that each $g_i$ has eigenvalues $\pm 1$. With this, we define a physical state as a state $\ket{\psi}$ that is unchanged by all local gauge transformations:
\begin{align}
    g_i\ket{\psi} = \ket{\psi} \; \forall i
\end{align}
At this point we will note that in previous sections we described our algorithm using a $G^2$ operator (defined in equation (10)), but this was for systems governed by compact Lie gauge groups. Here, we have the discrete group $\mathbb{Z}_2$, which requires two small alterations. Firstly, we replace the $G^2$ operator with a new operator $G_{tot}$:
\begin{align}
    G_{tot} = g_1 + g_2 + g_3 + g_4
\end{align}
We note that the spectrum of $G_{tot}$ is $-4$,$0$, and $4$; furthermore, using equations (19) and (20), we have that
\begin{align}
    G_{tot}\ket{\psi} = 4\ket{\psi}
\end{align}
for all physical states. This $G_{tot}$ operator serves the same purpose as the $G^2$ operator: every physical state is an eigenstate with the same eigenvalue (4 for $G_{tot}$ and 0 for $G^2$) and all unphysical states have different eigenvalues. Thus, $G_{tot}$ will function the same in our algorithm for our discrete group as $G^2$ will for compact Lie gauge groups.

Here, we must highlight the second alteration needed for our discrete group. Figure \ref{fig:Rodeo} shows phase rotation gates as part of the circuit diagram, but we argued that they are not required in our algorithm in section IIA due to physical states having an eigenvalue of 0. As just discussed, this is not true for our toy model, so we must include the phase rotation gates with $E=4$, the eigenvalue of physical states for our system. 

Now, our aim is to study gauge drift, so we include in our toy model an artificial gauge drift operator $D$ with each time step. We base our constructed gauge drift operators off of eq. (\ref{eq:drift}); specifically, we connect each of the, in our case 32, physical states of the system to a randomly selected unphysical state in exactly the way laid out in eq. (\ref{eq:drift}). All other states are unchanged. This provides a way for the system to drift out of the physical subspace of its Hilbert space in a unitary way, and a parameter, $\epsilon$, to control the strength of this drift. For our simulations, we set this parameter to $\epsilon = 0.01$. Note that each time our method calls for time evolution, we select a different set of 32 unphysical states to appear in the gauge drift operator so as to not bias any unphysical states. 

With this defined, the algorithm can be used for the toy model. Each time step takes a state $\ket{\psi}$ to $\tilde{g}_iDU_t\ket{\psi}$ where $\tilde{g}_i$ is a randomly chosen gauge transformation, $D$ is a gauge drift operator, and $U_t = \exp{-iHt}$ is the standard time evolution operator. Note that $\tilde{g}_i$ represents any gauge transformation on the system, including the $g_i$'s in equations (15)-(18) as well as any product of them. The remaining part of the algorithm is the rodeo projection, which occurs after a set number of time steps (determined in the calibration phase). As discussed above, a successful projection takes a state $\ket{\psi}$ to $\frac{1}{2}(1+\exp(-i(G_{tot} - 4)t_r))\ket{\psi}$ where here we note that the time used within the projection $t_r$ is in general different from the time step size $t$. Recall that all physical states are eigenstates of $G_{tot}$ with an eigenvalue of 4, so they are unaffected while unphysical states will suffer destructive interference. 

\begin{figure}[h]
    \centering
    \includegraphics[width=\linewidth]{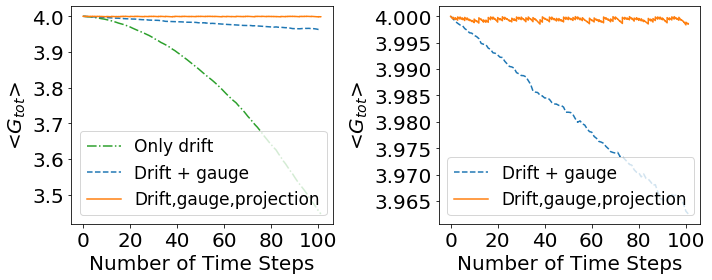}
    \caption{Comparing gauge drift suppression methods, including no suppression methods (dash-dot line), a random gauge transformation after each time step (dashed line), and a random gauge transformation followed by a projection after each time step (solid line). The right plot is the same as the left plot without the dash-dot plot so as to zoom in on the behavior of the other two.}
    \label{fig:Drift}
\end{figure}

Figure \ref{fig:Drift} demonstrates the basic performance of the method. It shows no suppression, a method of using a random gauge transformation after each time step, and our method of using a random gauge transformation after each time step plus projections, in this case after every time step. It can be seen both that the use of random gauge transformations after every time step greatly suppresses the gauge drift as well as that frequent rodeo projections does in fact keep the system very close to the physical subspace. 

\begin{figure}[h]
    \centering
    \includegraphics[width=\linewidth]{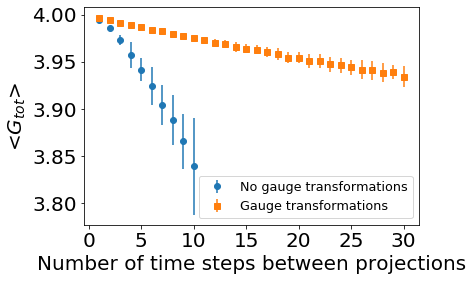}
    \caption{Comparing performance of regimes with no gauge transformations and various frequencies of projections (circle markers) with regimes with gauge transformations and various frequencies of projections (square markers)}
    \label{fig:DRvsDGR}
\end{figure}

In the rest of this section we will consider the frequency of  projections. To study this, we have run simulations to test many different regimes to compare their effectiveness. In particular,   systems without any gauge transformations after time steps with a frequency of rodeo projections ranging from after one time step to after 10 time steps. We also considered systems with a random gauge transformation after each time step with a frequency of gauge projections ranging from after one time step to after 30 time steps. For each regime, we ran our algorithm on 200 systems and calculated the average of the farthest from physical each run reached (which is to say the lowest value of $\expval{G_{tot}}$). 

Figure \ref{fig:DRvsDGR} shows these results.  It is clear that  including the random gauge transformations into the method significantly helps to reduce the required frequency of gauge projections. In particular, it shows the following approximate equivalences: a  projection after every time step performs similarly to a projection every two time steps when gauge transformations are included; a rodeo projection after every two time steps performs similarly to a gauge projection every five time steps when gauge transformations are included; and a gauge projection after every three time steps performs similarly to a gauge projection every ten time steps when gauge transformations are included. This shows that including the gauge transformations reduces the required frequency of rodeo projections by more than half. In particular, we see that for a system that drifts out of the physical subspace slowly, one could perform projections infrequently and still be able to utilize the Zeno effect to great advantage.

\section{Implementation}
It is important to consider the implementation of this method, as well as the associated costs. As our method involves two main operations (rodeo projections and gauge transformations), we will take them in turn. 

\subsection{Implementing Rodeo Projections}
With a circuit diagram of Fig.~\ref{fig:Rodeo} plus the small alterations discussed in section IIA in hand, there remain two main questions when it comes to the implementation of our version of rodeo projections: how should the time parameters, call them $t_r$, be chosen, and how many ancillary qubits per projection are required?

Firstly, we address the issue of the selection of the $t_r$ parameters. Here, we note that previous work has shown that the rodeo algorithm can be optimized and made far more efficient \cite{RodeoOpt} via considered choices for the parameter $t_r$. We suggest that this optimized approach makes sense for our purposes as well. This work notes that the dominant cost of the rodeo algorithm is the controlled time evolution gate. In our case it is then the controlled U gate with $U = \exp(-iG^2t_r)$. We will also note that constructing $G^2$ will be much more difficult than $H$ so our version of a rodeo projection suffers a much heftier cost in construction of the required gates, but so will alternative projection schemes. Due to the controlled U gate dominating costs, the authors of \cite{RodeoOpt} argue that the total time to conduct a rodeo projection acts as a proxy for computational costs and then proceeds to lay out a method for selecting the $t_r$s that maximizes efficiency of the algorithm given fixed computational resources \cite{RodeoOpt}. The authors note that choosing the $t_r$s randomly (eg via Gaussian distributions) can lead to exponentially large fluctuations in the suppression of undesired amplitudes; furthermore, they show that considered choices of the $t_r$s, specifically choosing $t_r$s that vary across exponentially many timescales, can eliminate the problem of these large fluctuations, thus leading to a more efficient implementation of the rodeo projection.

Additionally, we must address the question of qubit costs.
Notably, if mid-circuit measurements are allowed, then only one ancillary qubit is required, as it can be reused again and again. If this is not the case, then we estimate one projection would need roughly as many ancillary qubits as sites in the lattice. Future work aims to study how the qubit per projection requirement scales with system size.

\subsection{Implementing Gauge Transformations}
It is important to discuss the practical implementation of random gauge transformations as \cite{LLY} outlines in section III of their paper. First, we must sample from the Lie group $G$ of gauge transformations. One way is to construct an ancillary 'clock' register that increases its value by 1 after each gauge transformation is conducted in the course of our algorithm. Then a random circuit (eg of the kind developed in \cite{Short}) will take the value of the clock and select a gauge transformation for each site and store it on an ancillary G-register. Once sampled, these gauge transformations can be performed on the system.

The first major cost consideration of this process is qubit costs. The sampling process requires two additional registers (the clock and the G-registers); notably, this qubit cost does not scale with the size of the system. Furthermore, the actual implementation of the gauge transformations on the system "requires about as many G-multiplication gates as the original quantum simulation did" \cite{LLY}.

A second cost consideration is the depth of the random circuit. Intuitively, shorter circuits will result in less fair sampling from the Haaar measure of the Lie group than longer circuits. Luckily, we are helped significantly by the fact that we have no strict requirements on fair sampling from the Haar measure. Here, we point to a common method of sampling uniformly from the Haar measure of a group $G$, which is to select a small number (eg 2) of elements of $G$, call them $g_1$ and $g_2$, and construct strings of these elements of a chosen length (an example of a string of length 5 is $g_1g_2g_2g_1g_2$). Importantly, this method of sampling elements of $G$ converges to the Haar measure in the limit of long strings. This tells us that unfair sampling can be made more fair by conducting multiple unfairly sampled gauge transformations. Thus, we can use a short random circuit when sampling gauge transformations and then perform multiple gauge transformations in a row if need be to increase the fairness of the sampling. Future work aims to clarify the relationship between the fairness of sampling and the performance of the algorithm.

\section{Conclusion}

We have outlined an algorithm developed to maintain gauge invariance (or more precisely suppress gauge drift) during the simulation of lattice gauge theories. Our algorithm was based off of the Zeno effect, wherein we use frequent projections (i.e. altered rodeo projections) to keep the system within the physical subspace of its exponentially larger Hilbert space. We also include a technique of performing gauge transformations after every time step to help reduce the system's gauge drift into the unphysical sector which in turn allows for less frequent projections while still taking full advantage of the Zeno effect. We have shown that our method is effective for a simple small  toy model of pure $\mathbb{Z}_2$ gauge theory. Future work aims to test our algorithm out on larger systems, and more complicated gauge theories such as $SU(N)$ or gauge theories that include matter. 

\begin{acknowledgments}
This work was supported in part by the U.S. Department of Energy, Office of Nuclear Physics under Award Number(s) DE-SC0021143 and DE-FG02- 93ER40762.
\end{acknowledgments}

%\nocite{*}
\bibliography{biblio.bib}
\end{document}